\documentstyle[11pt]{article}
\catcode`\@=11
\begin{document} \openup6pt

\title{ Relativistic Models Of A Class Of Compact Objects
}

\author{ 
 Rumi Deb$^{1}$\thanks{Electronic mail : siya\_deb@yahoo.co.in} \\
 Bikash Chandra Paul$^{1,2}$\thanks{Electronic mail : bcpaul@iucaa.ernet.in} \\
   Ramesh Tikekar$^{3}$\thanks{Electronic mail : tikekar@iucaa.ernet.in} \\
   $^{1}$ IUCAA Reference Centre, Physics Department \\
North Bengal University, 
Dist. : Darjeeling, Pin : 734 013,  India \\
   $^{2}$Physics Department, North Bengal University, \\
Siliguri, Dist. : Darjeeling, Pin : 734 013,  India \\  
$^{3}$ Inter-University Centre for Astronomy and Astrophysics \\
P.O. Box 4, P.O.:Ganeshkhind, Pune, Pin: 411007, India 
}

\date{}

\maketitle

\vspace{0.5in}

\begin{abstract}

A class of  general relativistic solutions in isotropic spherical polar coordinates  are discussed which describe  compact stars  in hydrostatic equilibrium. The stellar models obtained here are characterized by four parameters, namely,  $\lambda$, $k$, $A$ and $R$ of geometrical significance related with inhomogenity of the matter content of the star. The stellar models obtained using the solutions  are  physically viable for a wide range of values of the parameters.  The  physical features of the compact objects taken up here are studied  numerically  for a number of admissible values of the parameters. Observational stellar mass data are used to construct suitable models of the compact stars. \\

\vspace{0.1 cm}

PACS No(s). 04.20.Jb, 04.40.Dg, 95.30.Sf \\


\end{abstract}

\vspace{0.2cm}


\vspace{4.5cm}

\pagebreak

\section{ Introduction:}

The discovery of compact stellar objects, such as X-ray pulsars, namely Her X1, millisecond pulsar SAX J1808.43658, X-ray sources 4U 1820-30 and 4U 1728-34 which are regarded as  the probable strange star candidates, has led to critical studies of relativistic models of such stellar configurations [1-10]. There are several such astrophysical as well as cosmological  situations where one needs to consider the equation  of state of matter involving matter densities of the order of  $10^{15} \;g \; cm^{-3}$ or higher, exceeding the nuclear density. The conventional approach of obtaining models of relativistic stars in equilibirium heavily relies on the availibility of definite information about the equation  of state of its matter content. Our knowledge about possible equation of state inside a superdense strange star at present is not known. In this context Vaidya-Tikekar \cite{vt} and Tikekar \cite{t} have shown that in the absence of definite information about equation  of state of matter content of stellar configuration, the alternative approach of prescribing suitable {\it ansatz}  geometry for the interior physical 3-space of the configuration leads to simple easily tractable models of such stars which are physically viable. Relativistic models of superdense stars based on different solutions of Einstein's field equations obtained by using Vaidya-Tikekar approach of assigning different geometries with physical 3-spaces of such objects have been studied by several workers \cite{mpd,th,tj,jt}. Pant and Sah \cite{ps} obtained a class of relativistic static non-singular analytic solutions in isotropic form describing space time of static  spherically symmetric  distribution of matter. The solution has been found to lead to a  physically viable causal model of neutron star with a maximum mass $4 M_{\odot}$. 

In this paper we discuss a class of  solution of relativistic field equations as obtained in Ref.   \cite{ps} and examine physical plausibility of several models of a class of neutron stars using numerical procedures to explore the possibility of using it to describe interior  of a compact star. It is also possible to estimate the radius of a star when its mass is known. It is also possible to determine the variation of matter density on its boundary surface and that at the center of a superdense star for the prescribed geometry.  
The plan of the paper is as follows :
in sec 2. the relevant relativistic field equations have been  set up and their solution is discussed. In sec 3. several features of physical relevance have been reported. In sec. 4, stellar models are discussed with the observational stellar mass data for different values of the parameters $\lambda$, $k$, $A$ and $R$.  Finally  in sec 5, we give a brief discussion.
 
\section{Field Equation and Solution }

The Einstein's field equation is
\begin{equation}
R_{\mu \nu} -\frac{1}{2} g_{\mu \nu} R =  8 \pi G  \; T_{\mu \nu}
\end{equation}
where  $g_{\mu \nu}$, $R$, $R_{\mu \nu}$ and $T_{\mu \nu}$ are the  metric tensor, Ricci scalar,  Ricci tensor and energy
momentum tensor respectively.
We use the following form of the space time metric given by
\begin{equation}
ds^2= e^{\nu(r)} dt^2-e^{\mu(r)}(dr^2+r^2d\Omega^2)
\end{equation}
with
\begin{equation}
d\Omega^2=d\theta^2+ sin^2\theta \; d\phi^2. 
\end{equation}
using isotropic spherical polar coordinate.
 In the next section  we use systems of units with $8\pi G=1$, $c=1$ respectively.

The energy momentum tensor for a spherical distribution of matter in the form of perfect fluid in equilibrium is given by
\begin{equation}
T^{\mu}_{\mu} = diag \; ( \rho, -p,
-p, -p)
\end{equation}
 where  $\rho$ and $p$ are energy density and fluid pressure of matter
respectively.  Using the space time metric given by eq.(2), the Einstein's field eq. (1) gives the following equations : 
\begin{equation}
\rho=-e^{-\mu}\left(\mu''+\frac{\mu'^2}{4}+\frac{2\mu'}{r}  \right)
\end{equation}
\begin{equation}
p=e^{-\mu}\left(\frac{\mu'^2}{4}+\frac{\mu'}{r}+\frac{\mu'\nu'}{2}+\frac{\nu'}{r}\right)  
\end{equation}
\begin{equation}
p=e^{-\mu}\left(\frac{\mu''}{2}+\frac{\nu''}{2}+\frac{\nu'^2}{4}+\frac{\mu'}{2r}+\frac{\nu'}{2r}\right).  
\end{equation}
 Now, pressure isotropy condition from eqs.(6) and (7) leads to the  following relation between metric variables $\mu$ and $\nu$:
\begin{equation}
\nu''+\mu''+\frac{\nu'^2}{2}-\frac{\mu'^2}{2}-\mu'\nu'-\frac{1}{r}\left(\nu'+\mu'\right)=0
\end{equation}
It is a  second order differential equation which permits a solution \cite{ps} as follows :
\begin{equation}
e^{\frac{\nu}{2}}=A \left(\frac{1-k\alpha}{1+k\alpha } \right), \;\;\; \;\;\;  e^{\frac{\mu}{2}}=\frac{(1+k\alpha)^2}{1+\frac{r^2}{R^2}}
\end{equation}
where $R$, $\lambda$, $k$ and $A$ are arbitrary constants. In the above we denote
\begin{equation}
\alpha(r)=\sqrt{\frac{1+\frac{r^2}{R^2}}{1+ \lambda \frac{ r^2}{R^2}}}.
\end{equation}
We observe that the geometry of that of the 3-space with metric
\begin{equation}
d\sigma^2=\frac{dr^2+r^2(d\theta^2+sin^2\theta d\phi^2)}{1+\frac{r^2}{R^2}}
\end{equation}
is that of a 3 sphere immersed in a 4-dimensional Euclidean space. Accordingly the geometry of physical space obtained at the $t=constant$  section of the space time is given by
\begin{equation}
ds^2=A^2\frac{(1-k\alpha)^2}{(1+k\alpha)^2}dt^2-\frac{(1+k\alpha)^4}{1+\frac{r^2}{R^2}}(dr^2+r^2(d\theta^2+sin^2\theta d\phi^2))
\end{equation}
where, $\alpha(r)$ is given by eq.(10). Hence the geometry of the 3 space obtained at $t=constant$ section of the space time of metric (12) is a deviation introduced in spherical 3 space and the parameter $k$ is a geometrical parameter measuring inhomogenity of the physical space. With $k=0$, the space time metric (12) degenerates into that of Einstein's static universe which is filled with matter of uniform density. The space time metric of Pant and Sah \cite{ps}  is a generalization of the Buchdahl solution, the physical 3-space associated with which has the same feature. 
For $\lambda=0$, the solution reduces to that obtained by Buchdahl which is an analog of a classical polytrope of index 5. However, for $\lambda>0$, the solution corresponds to finite boundary models. Pant and Sah \cite{ps} obtained a class of non-singular analytic solution of the general relativistic field equations for a static spherically symmetric material distribution which is matched with Schwarzschild's empty space time. In this paper we study physical properties of compact objects taking different values of  $R$, $\lambda$, $k$ and $A$ as permitted by the field equations. 
Using solution given by eq.(9) in eqs.(5)-(7), one obtains the explicit expressions  for the energy density and fluid pressure as follows:
\begin{equation}
\rho=\frac{12(1+\lambda k\alpha^5)}{R^2(1+k\alpha)^5},
\end{equation}
\begin{equation}
p=\frac{4(\lambda k^2\alpha^6-1)}{R^2(1+k\alpha)^5(1-k\alpha)}.
\end{equation}
The exterior Schwarzschild line element is given by
\begin{equation}
ds^2= \left( 1 - \frac{2m}{r_o} \right) dt^2 - \left( 1 - \frac{2m}{r_o} \right)^{-1} dr^2 - r_o^2 (d\theta^2 + sin^2 \theta d\phi^2)
\end{equation}
where $m$ represents the mass of spherical object. The above metric can be expressed in an isotropic form \cite{jvn}
\begin{equation}
ds^2=\left(\frac{1-\frac{m}{2r}}{1+\frac{m}{2r}}\right)^2dt^2-\left(1+\frac{m}{2r}\right)^4(dr^2+r^2d\Omega^2)
\end{equation}
using the transformation $r_o= r \left(1+ \frac{m}{2r} \right)^2$  where $r_o$ is the radius of the compact object. This form of the Schwarzschild metric will be used here to match at the boundary with the interior metric given by eq. (12).

\section{Physical properties of a compact star}
The solution given by eq.(9) is useful to study physical features of a compact star in a general way which are outlined as follows:

(1) In this model, $\rho$ and $p$ are determined using eqs.(13) and (14). We note that $\rho$ is obviously positive for any positive $\lambda$ and $k$, while $p\geq 0$ leads to two different cases: (i) $\lambda>1/k^2\alpha^6$ with $k<1/\alpha$ and (ii) $\lambda<1/k^2\alpha^6$ with $k>1/\alpha$.

(2) At the boundary of the star ($r=b$), the interior solution should be matched with the isotropic form of Schwarzschild exterior solution,i.e.,
\begin{equation}
e^{\frac{\nu}{2}}|_{r=b}=\left( \frac{1-\frac{m}{2b}}{1+\frac{m}{2b}}\right)\;\;\;\;\;\;\;  e^{\frac{\mu}{2}}|_{r=b}=\left(1+\frac{m}{2b}\right)^2
\end{equation}

(3) The physical radius of a star $r_o$, is determined  knowing the radial distance where  the pressure at the boundary vanishes (i.e., $p(r)=0$ at $r=b$). The  physical radius is related to the radial distance ($r=b$) through the relation $r_o= b \left(1+ \frac{m}{2b} \right)^2$ \cite{jvn}.

(4) the ratio $\frac{m}{b}$ is determined using eqs. (9) and (16), which is given by 
\begin{equation}
\frac{m}{b} = 2 \left( \frac{1+k \alpha}{\sqrt{1+y^2}} -1 \right)
\end{equation}

(5) The density inside the star should be positive i.e., $\rho>0$.

(6) Inside the star the stellar model should satisfy the condition, $dp/d\rho<1$ for the sound propagation to be causal.

The usual boundary conditions are that the first and second fundamental forms be continuous across the boundary $r=b$. Applying the boundary conditions we determine $A$ which is given by 

\begin{equation}
A=\frac{\left({1-\frac{m}{2b}}\right)}{\left({1+\frac{m}{2b}}\right)}
\left(   \frac{\sqrt{1+\lambda \frac{b^2}{R^2}} + k \sqrt{1+  \frac{b^2}{R^2}} }{ \sqrt{1+\lambda \frac{b^2}{R^2}} - k \sqrt{1+  \frac{b^2}{R^2} }}   \right)  
\end{equation}
Equating eqs.(9) and (16) at the boundary $(r=b)$, we get a eighth order polynomial equation in $y$ (here $\frac{b}{R}$ is replaced by $y$):
\[
[ (1+A)^4+k^4(1-A)^4-8(1+A)^2+16-2k^2(1-A^2)^2-8k^2(1-A)^2 ] +
[2\lambda(1+A)^4-16\lambda(1+A)^2
\]
\[
-8(1+A)^2+32(1+\lambda)-2k^2(1-A^2)^2(1+\lambda)-8k^2(2+\lambda)(1-A)^2+2k^4(1-A)^4]y^2
\]
\[
+
[\lambda^2(1+A)^4-8\lambda^2(1+A)^2-8\lambda(1+A)^2+(1+4\lambda+\lambda^2)-2\lambda k^2(1-A^2)^2
\]
\begin{equation}
-8k^2(1-A)^2(1+2\lambda)+k^4(1-A)^4]y^4-
[8\lambda^2(1+A)^2-32(1+\lambda)-8\lambda k^2(1-A)^2]y^6+16\lambda^2 y^8=0
\end{equation}
where $\lambda$, $k$ and $A$ are constants.
Imposing the condition that pressure at the boundary vanishes in eq.(14), we determine $y$  which is given by,
\begin{equation}
y=\sqrt{\frac{1-\left(\lambda k^2\right)^{1/3}}{\left(\lambda k^2\right)^{1/3}-\lambda}}.
\end{equation}
Thus, the size of a star is determined by  $k$ and $\lambda$. It is evident that a real   $y$ is permitted when (i) $k>\lambda$ with $\lambda<1$,   or (ii) $k<\lambda$ with $\lambda>1$. Using eqs.(20) and (21), a polynomial equation in $\lambda$, $k$ and $A$ is obtained.  
 Although the eq.(20) is a polynomial of degree eight  we note that only one realistic solution for $y$ is obtained for different domains of the values of any pair of parameters namely,  $A$, $k$  and $\lambda$. Subsequently the other parameters may be determined.
 For example, (i)  when $A=2$, we found that  $\lambda$ and   $k$ satisfy the following  inequalities $2.9\leq k \leq5$ and $1.4877\times 10^{-6}\leq \lambda \leq0.04$, (ii) when $A=4$, the range of permitted values are $1.7\leq k \leq 2.3$ and $0.0185 \leq \lambda\leq 0.0653$.  However, for a given $\lambda$, e.g.,
(i) $\lambda=0.15$, we note that the permitted values of $A$ lies in the range $3.6< A <5.6$, and (ii)that for  $\lambda=0.1318$, one obtains realistic solution for $3.5< A <5.8$.

The square of the acoustic velocity  $\frac{dp}{d\rho}$ takes the form : 
\begin{equation}
\frac{dp}{d\rho}=\frac{6 \lambda k \alpha^5(1-k \alpha)(1+k \alpha)-5(1-k \alpha)(\lambda k^2 \alpha^6-1)+(\lambda k^2 \alpha^6-1)(1+k \alpha)}{15(1-k \alpha)^2(\lambda \alpha^4(1+k \alpha)-(1+ \lambda k \alpha^5))}.
\end{equation}

A variation of $\frac{dp}{d\rho}$ for $\lambda = 0.1318$ and $k=2.2268$ is displayed in table 1. 
It is evident that $\frac{dp}{d\rho}$ is maximum at the center and gradually decreases outward. It is also found that inside the star the constrain  $\frac{dp}{d\rho}<1$ is always maintained which ensures causality. In table 2, variation of $\frac{dp}{d\rho}$  from the centre to the boundary for different values of $\lambda$ and $k$ are presented. It is evident that as $\lambda$ increases $\frac{dp}{d\rho}$ decreases at the centre. The variation of the central density with $\lambda$ and $k$ are displayed in tables (3) and (4) for $A=2$ and $A=4$ respectively. It is evident that the central density ($\rho_{c}$) decreases with an increase in $\lambda$. Thus stellar models with larger $\lambda$ accommodate a denser compact object compare to that for lower values of $\lambda$ and $k$. The  variation of pressure and density with radial distance are drawn employing eqs.(13) and (14) which are shown in figs.(1)-(4). Since it is not possible to express pressure in terms of density we study the behaviour of pressure and  density inside the curve numerically.In fig.(5) a variation of pressure with density is plotted for different model parameters.

\begin{table}[ht!]
\begin{center}
\begin{tabular}{|c|c|}  \hline
$r$ in the unit of $R$               & $\frac{dp}{d\rho}$     \\ \hline
0   & 0.521   \\ \hline
0.1   & 0.518   \\ \hline
0.2  & 0.513   \\ \hline
0.3   & 0.504   \\ \hline
0.4   & 0.496   \\ \hline
0.41   & 0.495   \\ \hline
0.42   & 0.495   \\ \hline
\end{tabular}
\caption{Variation of  $\frac{dp}{d\rho}$ with radial distance  $r$ for a given $\lambda=0.1318$ and $k=2.2268$}
\end{center}
\end{table}

\begin{table}[ht!]
\begin{center}
\begin{tabular}{|c|c|c|c|}  \hline
               & $\frac{dp}{d\rho}$  & $\frac{dp}{d\rho}$   & $\frac{dp}{d\rho}$        \\
               
$r$   & for  & for   & for   \\
in the unit of $R$ &  $\lambda=0.1211$ \& $k=2.2$   &   $\lambda=0.1318$ \& $k=2.2268$  &  $\lambda=0.15$\& $k=2.2681$
\\ \hline
0  & 0.524 & 0.521 & 0.520    \\ \hline
0.1  & 0.521 & 0.518 & 0.520    \\ \hline
0.2  & 0.514 & 0.513 & 0.513    \\ \hline
0.3  & 0.504 & 0.504 & 0.508    \\ \hline
0.4  & 0.494 & 0.496 &     \\ \hline
\end{tabular}
\caption{Variation of  $\frac{dp}{d\rho}$ with radial distance $r$ for different values of $\lambda$ and $k$.}
\end{center}
\end{table}

\begin{table}[ht!]
\begin{center}
\begin{tabular}{|c|c|c|}  \hline
$\lambda$               & $k$  & $\rho_c$ in the unit of $\frac{1.9 \times 10^{15}}{ R^{2}} \;  kg/m^3$    \\ \hline
$1.4877\times 10^{-6}$   & 2.9  & 0.0133  \\ \hline
$1.3836\times 10^{-5}$   & 3  & 0.0117   \\ \hline
0.0048   & 4  & 0.0039   \\ \hline
0.0400   & 5  & 0.0019   \\ \hline
\end{tabular}
\caption{Variation of  central density for $A=2$ for different values of $\lambda$ and $k$.}
\end{center}
\end{table}

\begin{table}[ht!]
\begin{center}
\begin{tabular}{|c|c|c|}  \hline
$\lambda$               & $k$  &$\rho_c$ in the unit of $\frac{1.9 \times 10^{15}}{ R^{2}} \;  kg/m^3$    \\ \hline
0.0185   & 1.7 & 0.0863   \\ \hline
0.0289  & 1.8  & 0.0734  \\ \hline
0.0432   & 1.9  & 0.0633  \\ \hline
0.0876   & 2.1  & 0.0496   \\ \hline
0.1211   & 2.2  & 0.0453   \\ \hline
0.15   & 2.268  & 0.0431   \\ \hline
\end{tabular}
\caption{Variation of  central density for $A=4$ for different values of $\lambda$ and $k$}
\end{center}
\end{table}

\input{epsf}
\begin{figure}
\epsffile{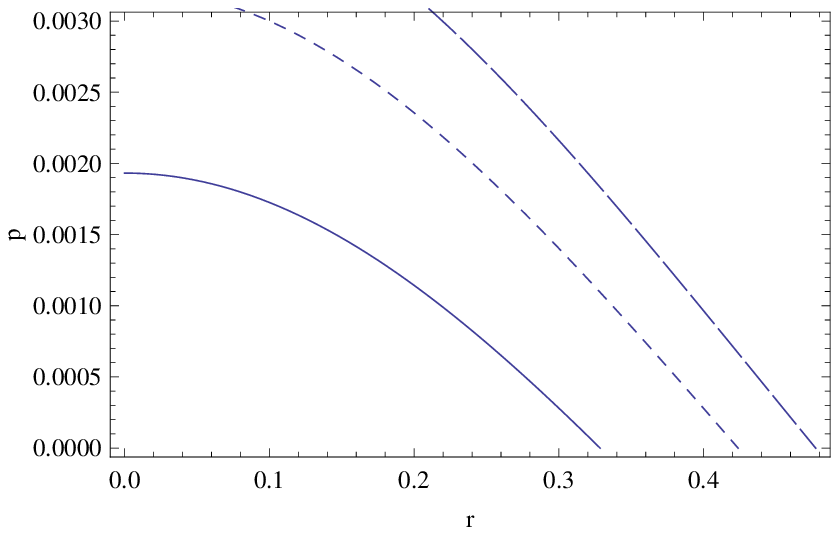}
\caption{  Variations of pressure with radial distance (in the unit of $R$) is plotted with solid line for $\lambda=0.15$, dashed line for $\lambda=0.1318$  and broken line for $\lambda=0.1211$.}
\end{figure}  

\input{epsf}
\begin{figure}
\epsffile{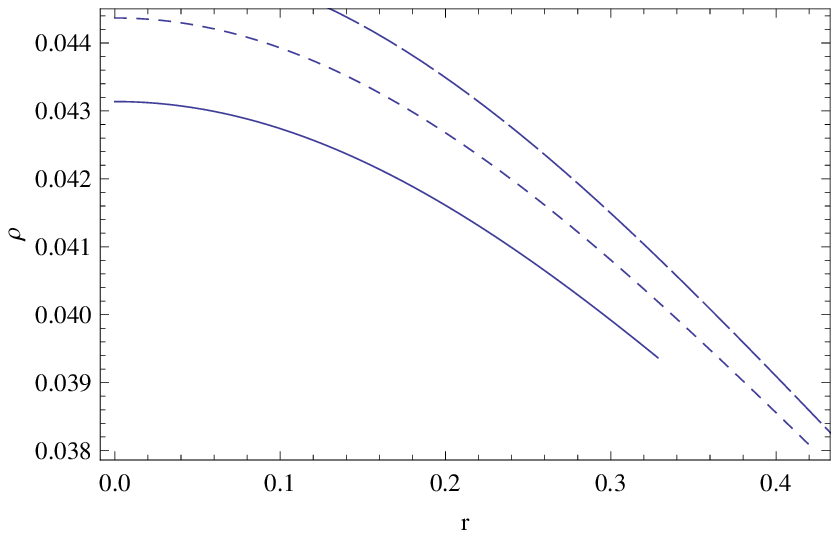}
\caption{ Variations of density with radial distance (in the unit of $R$ ) is plotted with solid line for $\lambda=0.15$, dashed line for $\lambda=0.1318$  and broken line for $\lambda=0.1211$.}
\end{figure}  

\input{epsf}
\begin{figure}
\epsffile{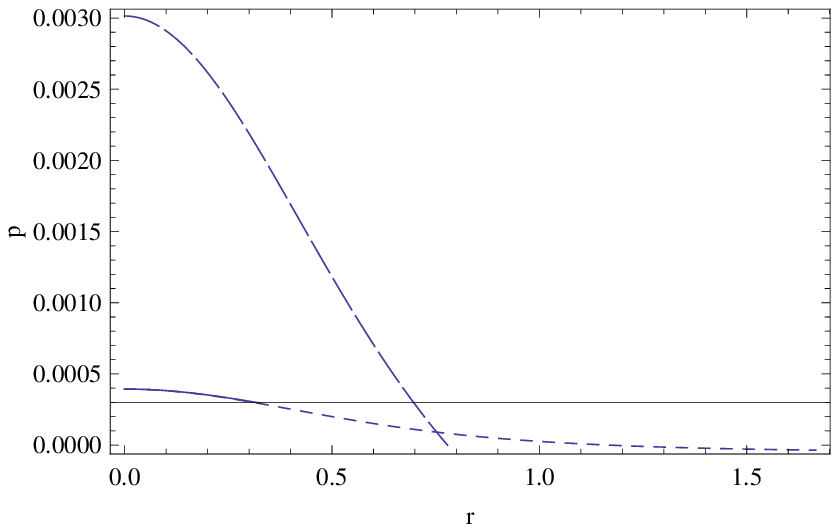}
\caption{  Variations of pressure with radial distance (in the unit of $R$ ) is plotted with solid line for $A=4$,  broken line for $A=3$ and dashed line for $A=2$.}
\end{figure}  

\input{epsf}
\begin{figure}
\epsffile{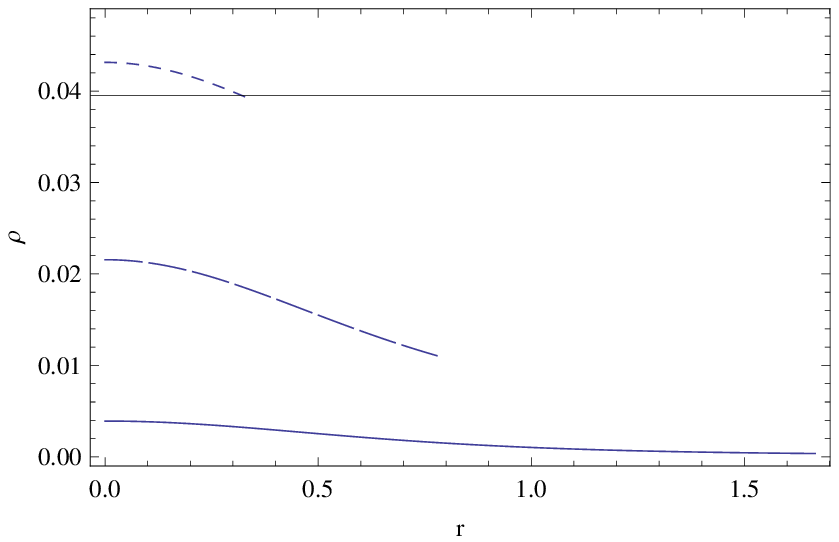}
\caption{  Variations of density with radial distance (in the unit of $R$ km ) is plotted with solid line for $A=4$,  broken line for $A=3$ and dashed line for $A=2$.}
\end{figure} 

\input{epsf}
\begin{figure}
\epsffile{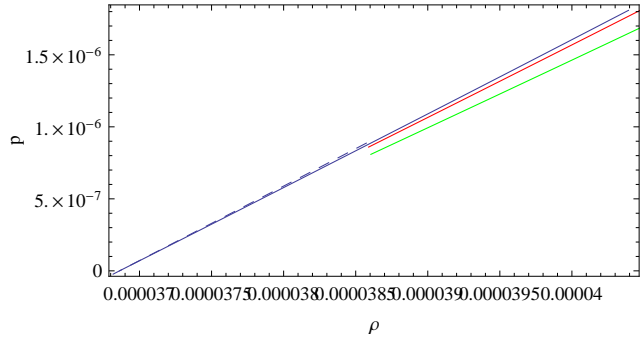}
\caption{  Variations of pressure   with density  is plotted with   green  line for $\lambda=0.0876$, red line for $\lambda=0.1318$, solid line for $\lambda=0.15$ and dahed line for $\lambda=0.165633$.}
\end{figure} 

\begin{table}[ht!]
\begin{center}
\begin{tabular}{|c|c|c|}  \hline

          $A=2$     & M (mass)  in $M_{\odot}$  & $r_o$ (radius) in km    \\ \hline
$\lambda=1.48\times 10^{-6}$, $k=2.9$   & 3.61 & 12.087   \\ \hline
$\lambda=1.17\times 10^{-5}$, $k=2.99$   & 2.69  & 9.250  \\ \hline
$\lambda=1.38\times 10^{-5}$, $k=3.0$   &  2.63 &  9.067   \\ \hline
$\lambda=3.93\times 10^{-2}$, $k=4.99$   & 0.12 &  0.622 \\ \hline
\end{tabular}
\caption{Variation of  mass and radius of a compact star for different values of $\lambda$ and $k$ for $R=0.2$ km.}
\end{center}
\end{table}

\begin{table}[ht!]
\begin{center}
\begin{tabular}{|c|c|c|}  \hline

       $A = 4$        & M (mass)  in $M_{\odot}$  & $b_o$ (radius) in km   \\ \hline

$\lambda=0.1211$, $k=2.2$   &  3.35 &  11.268   \\ \hline
$\lambda=0.1318$, $k=2.2268$   &  2.82 & 10.324  \\ \hline
$\lambda=0.15$, $k=2.2681$   & 2.45  &  8.409   \\ \hline
$\lambda=8.76\times 10^{-2}$, $k=2.1$   & 4.19  & 13.688   \\ \hline
$\lambda=0.1656 $, $k=2.3$   &  1.79 & 6.214    \\ \hline
\end{tabular}
\caption{Variation of  mass and radius of a compact star for different values of $\lambda$ and $k$ for  $R=2.5$ km.}
\end{center}
\end{table}

\begin{table}[ht!]
\begin{center}
\begin{tabular}{|c|c|c|c|c|c|c|}  \hline

               & $\lambda=1.48\times10^{-6}$  & $\lambda=8.76\times10^{-2}$    & $\lambda=0.1211$  & $\lambda=0.1318$  & $\lambda=0.15$  & $\lambda=0.1656$    \\ 
               &$k=2.9$  & $k=2.1$ & $k=2.2$  & $k=2.2268$  & $k=2.2681$   & $k=4.99$ \\ \hline
               
$\frac{\rho_b}{\rho_o}$ & $1.89\times10^{-5}$ & 0.45 & 0.54 & 0.58 & 0.69 & 0.81     \\ \hline
\end{tabular}
\caption{Variation of  the ratio of the density at the boundary to the density at the center of a compact star for different values of $\lambda$ and $k$.}
\end{center}
\end{table}

\section{Physical Analysis :}

In this section we analyze the physical properties of compact objects numerically. For given values of $\lambda$ and $k$, the radial coordinate at which the pressure vanishes may  be determined from eq.(14). The mass to radial distance $\frac{m}{b}$ is estimated from eq.(18), which in turn determines the physical size of the compact star ($r_o$).  For  a given set of values of the parameters $\lambda$, $A$ and  $k$, the mass ($m$) and radius of a compact object is obtained  in terms of the model parameter $R$. Thus for a known mass of a compact star $R$ is determined which in turn determines the corresponding radius. 
As the equation to determine the parameters  in the model is highly non-linear and intractable in known functional form, we adopt numerical technique in the next section.

 The radial variation of pressure and density of compact stars for   different parameters  are shown in figs. (1)-(5). It is evident that as  $\lambda$ is increased both the pressure and density at the centre is found to decrease and at the same time it corresponds to a smaller size  accommodating more mass.

  For a given  mass of a compact star \cite{tab},     it is possible to  estimate the corresponding radius in terms of the parameter $R$. We note that for a given mass of a compact star known from observation, the radius of the star may be estimated from a given $R$. However, as the radius of a neutron star is $\leq 10 \; km$, it is possible to obtain a class of stellar model taking different $R$ so that the size of the star is  satisfies the upper bound. In the next section we consider a few such stars whose masses are known from observations.
 
 We present below four different models using stellar mass data \cite{tab, dd, ddd} in the next section :

{\it Model 1 } :
 We consider X-ray pulsar Her X-1 \cite{tab, mdey, rs} which is characterized by  mass $M = 1.47 \; M_{\odot}$, where $M_{\odot}$  = the solar mass and found that it permits a star with radius $r_o
 = 4.921 $ km, for $R=0.081$ km. The compactness of the star in this case is  $ u=\frac{M}{r_o} = 0.30$.  The ratio of density at the boundary to that at the centre for the star is 0.0003 which is possible for the set of parameters $\lambda = 1.48 \times 10^{-6}$ and $k=2.9$. Taking different values of $R$  we get different models but a physically realistic model is obtained which accommodates a compact star with radius $\sim$ 10 km. For example, if $R= 2.504$ km, one obtains a compact object with radius $r_o=7.791$ km.
 In the later case we note that  the ratio of density at the boundary to that at the centre  is very high (0.99). The compactness of the star is 0.189 which is permitted for the set of parameters $\lambda = 0.0393$ and $k=4.99$ with  $A=2$.

{\it Model 2} : We consider X-ray pulsar 4U 1700- 37 which is characterized by mass $M = 2.44 \; M_{\odot}$ \cite{tab}. We note that for  $A=4$, $\lambda=0.1211$ and $k=2.2$, the corresponding radius of the above star is $ r_o = 8.197$  km with  $R= 1.819$ km. The ratio of density at the boundary to that at the centre for the star in this case is 0.820. However, for the set of values  $A=2$, $\lambda=0.1656$ and $k=2.3$, a compact object is permitted with radius $r_o = 8.110$ km  when $R=0.135$ km. The ratio of density at the boundary to that at the centre for the star in this case is 0.0003. Another stellar model is obtained for a set of values with $A=2$, $\lambda = 0.0393$ and $k=4.99$, where
the ratio of density at the boundary to that at the centre   is 0.99. In the later case the values is more compare to that one obtains taking $A=4$.  However both the cases permits a star with  compactness factor  $u=0.3$.

{\it Model 3} : We consider a neutron star J1518+4904 which is characterized by mass $M= 0.72 \; M_{\odot}$ \cite{tab}. For $\lambda =0.1211$, $k=2.2$ and $A=4$, the radius of the star estimated here is $r_o= 2.419 $ km with $R=0.537$ km. The ratio of density at the boundary to that at the centre for the star  is  0.82. In this case the compactness factor of the star is $u=0.3$. For $A=2$ we note the following : (i) when  $\lambda = 1.48 \times 10^{-6}$ and $k=2.9$, it admits  a star with radius 
 $r_o=2.4$ km for $R=0.04$ km and (ii) when $\lambda = 0.0393$ and $k=4.99$, it admits a star with radius for  $r_o=3.816$ km for  $R=1.226$ km.   The ratio of density at the boundary to that at the centre for the star in the first case is  0.0003 and that in the later case is 0.988. However, the compactness factor for the former is 0.3 which is higher than that in the  second case (0.189).

{\it Model 4} : We consider a neutron star J1748-2021 B  which is characterized by mass $M= 2.74 \; M_{\odot}$ \cite{tab}. For $A=4$, $\lambda = 0.1318$ and $k=2.2268$, a star of radius  $ r_o = 9.281$  km with $R=2.247$ km ids permmited . The ratio of density at the boundary to that at the centre for the star is 0.856. The  compactness factor is $u=0.3$. In the other case one obtains a star with  radius $ r_o = 8.467$  km with $R=3.406$ km when $\lambda = 0.1656$ and $k=2.3$. A star of smaller size is thus permitted in the later case with  compactness factor  (0.32) than that of the formal model.

 For $A=2$, stellar model admits a star with radius  $r_o=13.154 $ km for  $R=2.74 $ km, $\lambda = 0.138 \times 10^{-5}$ and $k=3$. However a smaller star with radius $r_o=8.380$ km is permitted here when  $R=0.181$, km with $\lambda = 1.17 \times 10^{-5}$ and $k=2.99$. The ratio of density at the boundary to that at the centre in the first case is 0.0017 which is higher than the later (0.0015). The compactness factor in the former model is 0.20 which is lesser than the later case 0.32.

\section{  Discussions : }

In this paper, we present general relativistic solution for a class of compact stars which are in hydrostatic equilibrium considering the isotropic form for a static spherically symmetric matter distribution. The general relativistic solution obtained by Pant and Sah \cite{ps} is employed here to study compact objects. We use isotropic form of the exterior Schwarzschild solution to match at the boundary of the compact object. The  stellar models  discussed here contains four   parameters $\lambda$, $A$,  $k$ and $R$. The observed mass of a star determines $R$ for known values of $\lambda$, $A$,  $k$.

We note the following:
(i) In fig. 1,  variation of pressure with radial distance is plotted for different  $\lambda$ for given values of  $A$ and $k$. The figures show that as $\lambda$ increases pressure decreases inside the  star.
(ii) In fig. 2, radial variation of density is plotted for different  $\lambda$. We note higher density for lower  $\lambda$. 
(iii) The variation of  $\frac{dp}{d\rho}$ inside the star for a given set of  values of $\lambda$  and $k$ are shown in table 1.  The causality condition is obeyed inside the star and $\frac{dp}{d\rho}$ is maximum at the center which however found to decrease monotonically radially outward. 
For different $\lambda$ and $k$, values of  $\frac{dp}{d\rho}$ is  also shown in table 2. It is evident that
$\frac{dp}{d\rho}$ decreases for an increase in $\lambda$ and $k$ values. 
(iv)  Variation of central density for different values of  $\lambda$ and $k$ with  $A=2$ and $A=4$  are presented separately in tables (3) and (4) respectively. We note that the central density decreases as the value for the pair ($\lambda$ and $k$) increases. From tables (3) and (4)  similar tendency for central density is found to exist when  $A$ is increased. As the isotropic Schwarzschild metric is  singular at $m=2b$, the model considered here may be useful to represent a strange star with $m \neq 2b$ or $m<2b$.
(v) In tables (5) and (6), the mass of a star with its maximum size is shown for different values of $\lambda$ and $k$ taking density of a star  $\rho_b=2\times 10^{15}\; gm/cc$ at the boundary. We obtain here a class of relativistic stars for different values of $\lambda$,  $A$, $k$ and $R$.
(vi) The density profile  of a given star with different values of $\lambda$ and $k$ is shown in table 7. As $\lambda$ increases the ratio of density at the boundary to that at the center is found to  increase accommodating more compact stars. 
(vii) In fig. 3, variation of pressure with radial distance is plotted for different values of  $A$. It is evident  that as $A$  increases  pressure  decreases. (viii) In fig. 4, variation of density with radial distance is plotted for different $A$. We note that as $A$ is increased both the density and the pressure decreases. But the  size of a star increases with an increase in $A$ thereby  accommodating more compact stars.
(ix) In fig. 5, variation of pressure with density is plotted for different $\lambda$.  We note that for a given density pressure is more for higher $\lambda$, this leads to a star with  higher central density. 

 In  sec. 4, we present models  of the  neutron stars that are tested for some known   compact objects. As the equation of state is not known we analyze the star for known geometry considered here. The radii of the compact stars namely, neutron stars are also  estimated here for known mass with a given $R$.  The parameter $R$ permits a class of compact objects, some of which are relevant observationally.  Considering observed masses of the compact objects  namely,  X-ray pulsars Her X-1,  4U 1700-37 and neutron stars J1518+4904, J1748-2021 B we analyze the interior of the star. We obtain a class of compact stars models for various $R$ with  given values of  $k$,  $\lambda$ and $A$. The stellar models obtained here can accomodate highly compact objects.  However  a detail study of the stellar composition at high pressure and density will be taken up elsewhere.

{\bf{ \it Acknowledgement :}}

BCP would like to acknowledge fruitfull discussion with Mira Dey and Jisnu Dey while visiting IUCAA, Pune. Authors would like to  thank IUCAA, Pune and IRC, Physics Department, North Bengal University (NBU) for providing facilities
to complete the work. BCP would like to thank  University Grants Commission, New Delhi for financial support. RT is thankfull to UGC for its award of Emiritus Fellowship. The authors would like to thank the referee for constructive criticism.

\pagebreak

\end{document}